\documentclass[reprint,superscriptaddress,prl]{revtex4-1}
\usepackage{graphicx}
\usepackage{amsmath}
\usepackage{amsfonts}
\usepackage{bm}
\usepackage{color}
\usepackage{array}
\usepackage{hyperref}
\hypersetup{%
 pdftitle = {Observation of guided acoustic waves in a human skull},
 pdfauthor = {H. Estrada},
 bookmarksopen = true,
 colorlinks  = true,
 citecolor   = black,
 linkcolor   = black,
 raiselinks  = true,
 urlcolor    = blue,
 pdfstartview= FitH,
}
\begin{document}
\title{Observation of guided acoustic waves in a human skull}
\author{H\'ector Estrada}
\email{Corresponding author: hector.estrada@posteo.org}
\affiliation{Institute of Biological and Medical Imaging (IBMI), Helmholtz Center Munich, Neuherberg, Germany}
\author{Sven Gottschalk}
\affiliation{Institute of Biological and Medical Imaging (IBMI), Helmholtz Center Munich, Neuherberg, Germany}
\author{Michael Reiss}
\affiliation{Institute of Biological and Medical Imaging (IBMI), Helmholtz Center Munich, Neuherberg, Germany}
\author{Volker Neuschmelting}
\affiliation{Institute of Biological and Medical Imaging (IBMI), Helmholtz Center Munich, Neuherberg, Germany}
\affiliation{Department of Neurosurgery, University Hospital Cologne, Cologne, Germany}
\author{Roland Goldbrunner}
\affiliation{Department of Neurosurgery, University Hospital Cologne, Cologne, Germany}
\author{Daniel Razansky}
\email{Corresponding author: dr@tum.de}
\affiliation{Institute of Biological and Medical Imaging (IBMI), Helmholtz Center Munich, Neuherberg, Germany}
\affiliation{Faculty of Medicine, Technical University of Munich, 81675 Munich, Germany}

\date{\today}
\begin{abstract}
Human skull poses a significant barrier for the propagation of ultrasound waves. Development of methods enabling more efficient ultrasound transmission into and from the brain is therefore critical for the advancement of ultrasound-mediated transcranial imaging or actuation techniques. We report on the first observation of guided acoustic waves in the near-field of an \textit{ex vivo} human skull specimen in the frequency range between 0.2 and 1.5 MHz. In contrast to what was previously observed for the guided wave propagation in thin rodent skulls, the guided wave observed in a higher frequency regime corresponds to a quasi-Rayleigh wave, mostly confined to the cortical bone layer. The newly discovered near-field properties of the human skull are expected to facilitate the development of more efficient diagnostic and therapeutic techniques based on transcranial ultrasound.
\end{abstract}

\maketitle
\section{Introduction}

Bones carry out important mechanical and hematopoietic functions and their mechano-structural anomalies, such as osteoporosis, can in principle be detected using ultrasonic methods \cite{boneUltrasound}. Yet, the human skull bone poses a challenge in the study of the brain by ultrasound-mediated techniques \cite{Fry1978}. Understanding the interaction of ultrasound waves with the human skull \cite{Fry1978, ClemeHUiM&B2002, Clement2002, ClemeWHTJotASoA2004, MarquPAMMTFPiMaB2009, PichaSHPiMaB2011, pinton2012} has been paramount in achieving focused ultrasound therapy deep inside human brain \cite{DanieWJoTU2014}. In small animals, optoacoustic neuroimaging techniques \cite{yw2014,Dean-GMSRCSR2017} have been successful in delivering transcranial images of cerebral vasculature by means of ultrasonic waves generated upon absorption of nanosecond laser pulses in the brain \cite{Estrada2016, KneipTERSRJoB2016}. While for high intensity focused ultrasound (HIFU) therapy the sources of narrowband ultrasound vibrations are located outside the head and far away from the skull, in optoacoustic neuroimaging applications the broadband ultrasound waves are mainly generated inside the brain in close proximity to the skull, supporting the existence of skull-guided acoustic waves (GAW) in mice \cite{EstraRRPiMaB2017}. GAW also exist in long cylindrical bones \cite{MoilaIToUFaFC2008,TalmaFM2011,MoilaZKKKPMHTUiMB2014}, enabling the assessment of cortical bone thickness and stiffness \cite{BochuVMLSR2017}. However, the inner structure of the human skull, composed of two cortical bone layers separated by a substantial layer of trabecular spongious bone (the diplo\"e), is considerably different from other types of bones. It also significantly deviates from the structure found in murine skulls \cite{BrookR1998,JilkaTJoGSA2013,EstraRRPiMaB2017,Estrada2016}, where the cortical bone layer occupies a larger proportion of the cross section while, in some regions, the diplo\"e is considerably thinner or non-existent. 

In addition to transcranial ultrasonography \cite{BaykoBMNRSSAP2003,LindsNBLSUiMB2014,ShapoSWMSMIToBE2015}, new ultrasound-mediated techniques are currently being proposed \cite{SeoNSSARCMN2016,SeoCRAMAe2013} to monitor neural activity in cortical areas of the human brain transcranially. It is thus desirable to extend the current knowledge about the near-field ultrasound wave propagation of the human skull.
  
\begin{figure}[h!]
 \includegraphics[width=\columnwidth]{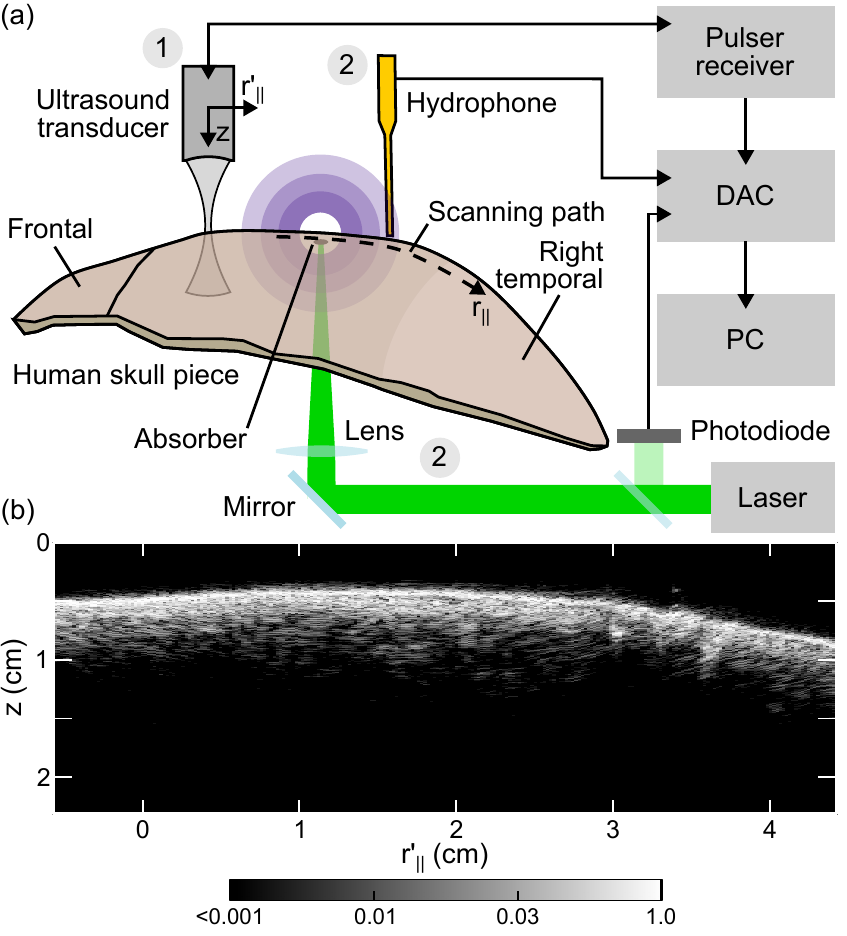}
 \caption{\label{fig:Diagram} \textbf{Experimental setup and skull surface mapping}. (a) Schematic of the experimental setup showing the water-immersed piece of human skull sample. The skull's surface is first extracted from a pulse-echo ultrasound scan (1). The propagation of waves generated by laser excitation of an optical absorber placed on the inner side of the skull is measured by a needle hydrophone scanned in close proximity (near-field) of the skull (2).The signals digitised by the acquisition card (DAC) are stored in a personal computer (PC), which also controls the scanning stages. (b) B-mode ultrasound scan of the skull’s cross section along the hydrophone scanning path.}
\end{figure}
\section{Materials and methods}
\subsection{Experimental setup}
Fresh frozen frontotemporal bone sample derived from decompressive hemicraniectomy was collected according to protocols established by the ethical committee of the Department of Neurosurgery at the University Hospital Cologne. The skull sample was kept at -80$^{\circ}$C and was later degassed for 3 hours and immersed in 0.9 \% saline solution during the imaging experiments that were performed in full compliance with the institutional guidelines of the Helmholtz Center Munich. 
To investigate on the exstence of GAW in a human skull, we adapted a similar approach to \cite{EstraRRPiMaB2017}, where the exact shape of the skull surface was first obtained from a pulse-echo scan (Fig. 1(b)) performed by a focused transducer (20 mm focal distance, 15 MHz central frequency, Olympus) with a step size of 200 $\mu$m. In order to excite broadband optoacoustic responses, a 200 $\mu$m thick layer of a black burnish attached on the interior side of the skull sample was illuminated with a 1 mJ laser pulses of 10 ns duration and 532 nm wavelength (Spectra Physics, USA) focused down to a 1 mm spot. The generated responses were recorded by a needle hydrophone (0.5 mm diameter, Precision Acoustics, UK) that was scanned following the path $r_\parallel$ in close proximity to the skull’s surface across the right temporal bone (see Fig. 1). The scanning step size (0.4 mm) was selected to allow the analysis in the spatial frequency domain up to 1.25 (mm$^{-1}$). The laser pulse was fired at a repetition frequency of 15 Hz and the data acquisition was synchronized using a photodiode (DET10A, Thorlabs, USA), to avoid jitter in the laser trigger signal (Fig. 1). The laser energy fluctuations were further accounted for using pulse-to-pulse photodiode measurements. The data was digitised at 60 MS$/$s by the data acquisition card (M3i.4142, Spectrum Systementwicklung Microelectronic, Germany) and stored for further analysis on a personal computer. 
\subsection{Simulations}
As a first approximation, we modeled a flat multilayered viscoelastic solid embedded in a fluid \cite{EstraRRPiMaB2017} by means of the global matrix method \cite{LoweUFaFCITo1995}. For a given frequency and wavevector $k_{\parallel}$ (incidence angle), a plane (inhomogeneous) wave is propagated from the input fluid, through the solid layers, to the output fluid. In the solid layers, longitudinal and transverse waves are considered in the propagation, as well as reflections at each interface between different media \cite{EstraRRPiMaB2017}, forming a linear system of equations with 14 unknowns (complex transmitted and reflected amplitudes in each medium). First, the transmission problem was solved at a given region of the reciprocal space (frequency-wavevector) and then the transmission maxima at the output fluid were extracted. Second, a modal solution of the system was found by further refining the position of the extracted transmission maxima in reciprocal space using a golden-section search for singularity of the global matrix. The calculation of the linear system of equations \cite{LoweUFaFCITo1995} was implemented in C++ and the analysis of the results was performed in Python. We assumed a total skull thickness of $h = 6$ mm (manually measured average), and elastic constants close to what has been reported in the literature for cortical and trabecular bones \cite{cgts2010} (see Table 1).

\begin{table}
\caption{Simulation parameters}
\bgroup
\def\arraystretch{1.4}%
\begin{tabular}%
  {>{\raggedright\arraybackslash}p{32mm}%
   >{\raggedleft\arraybackslash}p{12mm}%
   >{\raggedleft\arraybackslash}p{17mm}%
   >{\raggedleft\arraybackslash}p{15mm}}
 & Saline solution & Cortical bone & Diplo\"e\\\hline
Thickness (mm) & & (outer) 1.56 (inner) 1.44  & 3\\
Density (kg/m$^3$) & 1000 & 1969 & 1055\\
Longitudinal wave \newline speed (m/s) & 1504 & 3476 & 1886\\
Transverse wave \newline speed (m/s) & 0 & 1520 & 830\\ 
Volumetric viscosity \newline (Pa s) & 0 & 0.1 & 1.5\\
Shear viscosity \newline (Pa s) & 0 & 1 & 3
\end{tabular}
\egroup
\end{table}

\begin{figure}[t]
 \includegraphics[width=\columnwidth]{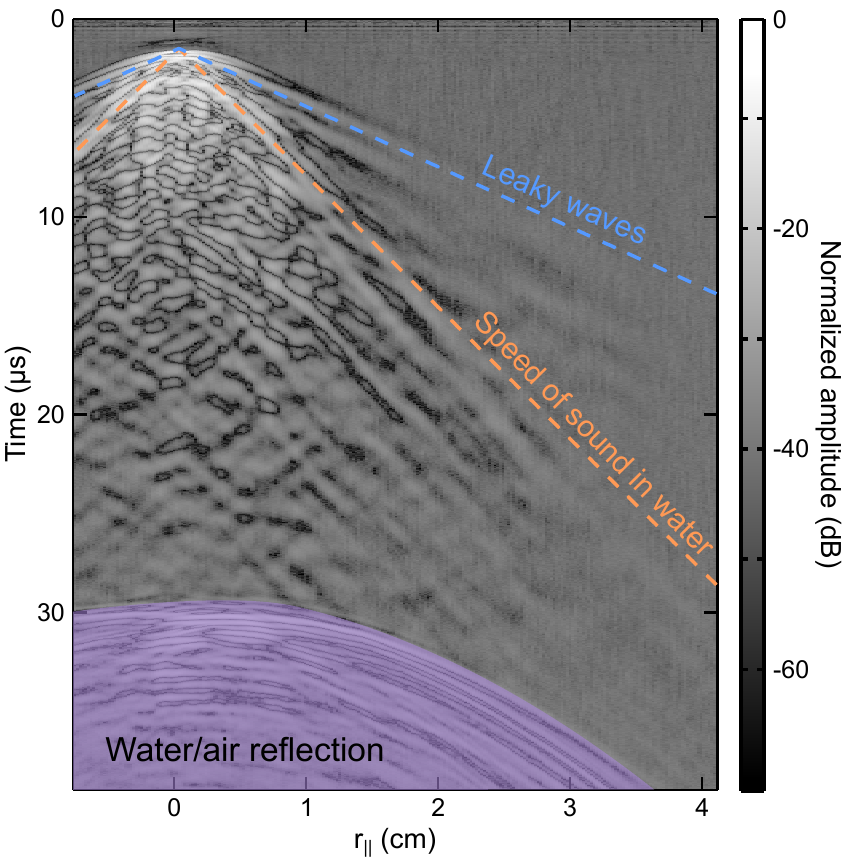}
 \caption{\label{fig:timeSpace} \textbf{Ultrasound wave propagation at the near-field of a human skull sample.} Normalized amplitude (gray scale in dB) of the signal detected by the needle hydrophone as a function of the time of flight and the measurement position relative to the optoacoustic source. The time delay from the initial laser pulse corresponds to the transit time of the elastic waves through the skull bone.}
\end{figure}%
\section{Results and discussion}
The amplitude of the detected signal is shown in Fig. 2 as a function of the distance to the optoacoustic source $r_{||}$ and the time of flight. The main wavefront is not registered at $t = 0$ because the source is placed on the opposite side of the skull (see Fig. 1(a)) and requires a finite amount of time to propagate to the hydrophone. Waves propagating at a speed faster than the speed of sound in water ($c_0$) can be clearly identified, similarly to the previously reported murine skull measurements \cite{EstraRRPiMaB2017}. 
The waves propagating with sonic speed are also distinguishable. However, the regime where waves propagate slower than $c_0$ is obscured by many oscillations that could be attributed to scattering events due to skull inhomogeneities or internal reverberations. To gain additional insight on the different simultaneously occurring ultrasound propagation regimes, one may further calculate the modes’ dispersion by means of a two-dimensional Fourier transform of the measured spatio-temporal data (Fig. 3). In the reciprocal space representation, the modes propagating with the speed of sound in water follow the green line. All the modes located above the sound line are therefore leaky while the ones below are non-leaky according to the Snell law. Several non-leaky modes are clearly distinguishable, in good agreement with the predictions of a flat multilayered solid model \cite{EstraRRPiMaB2017,LoweUFaFCITo1995}, which are labeled by solid white curves in Fig. 3. The modes follow an asymptotic trend propagating at 1343 (m$/$s) for frequencies higher than 0.5 MHz (see Fig. 3(b)).
We further evaluated the influence of first order corrections due to the skull's curvature following methodology described in \cite{SAWInhomo1995}. However, those turned negligible ($\simeq$-0.4 \%) due to the large radius of curvature of the skull ($\simeq$10 cm) compared to the acoustic wavelength $\lambda$ and the skull's thickness $h$ ($\simeq$ 6 mm).
The skull-guided waves in the case of murine skull were measured in the $0.3 < h/\lambda< 1.1$ range, whereas, due to constrains of the experimental methodology and the sample geometry, the current results for the human skull are shown in a different regime, i.e. $1.2 < h/\lambda<6.3$. 

\begin{figure}[t]
 \includegraphics[width=\columnwidth]{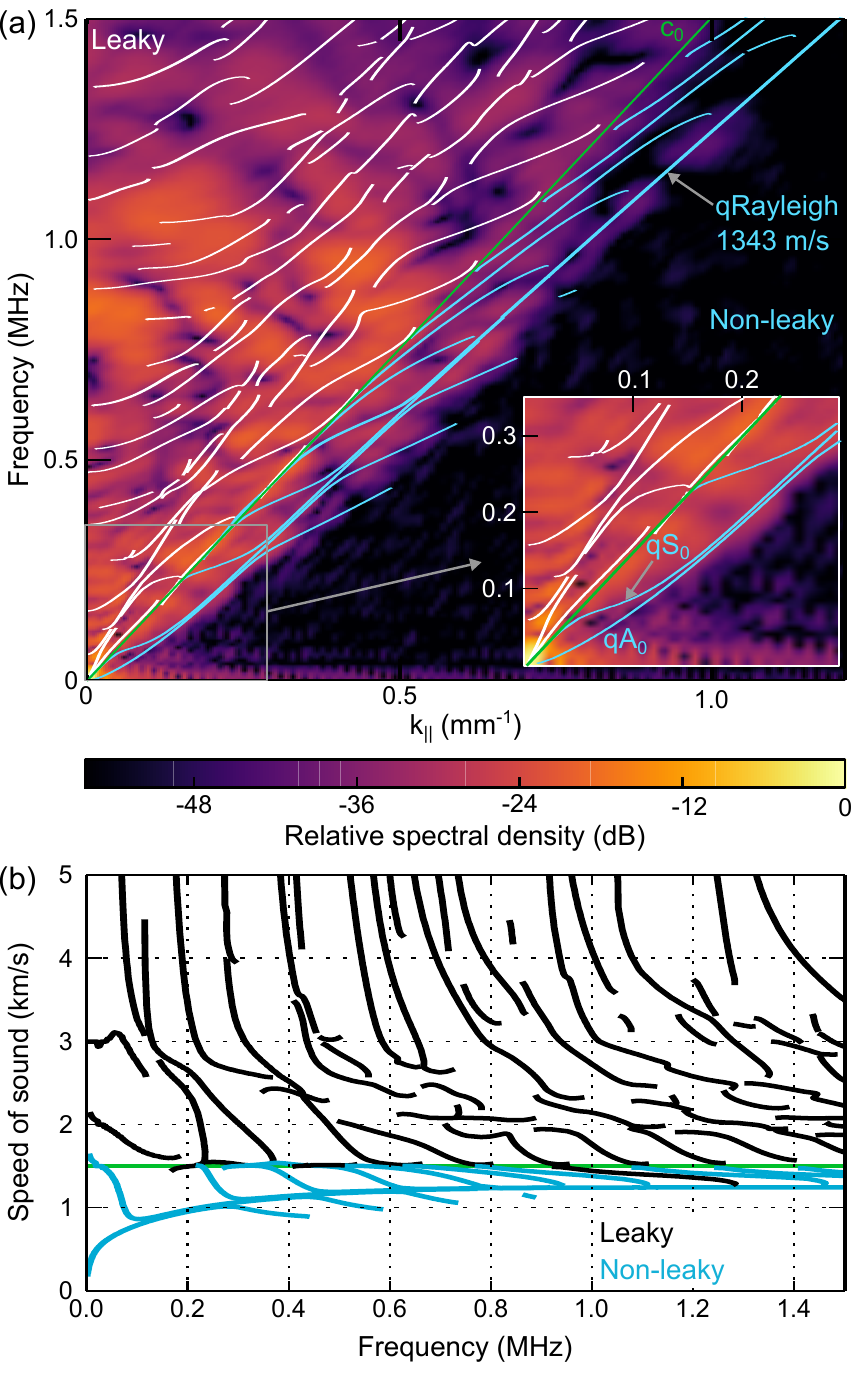}
 \caption{\label{fig:dispersion} \textbf{Modes' dispersion and speed of sound}. (a) Measured versus calculated (overlaid white and light blue lines) dispersion. The inset at the bottom right corner shows the details of the low frequency region. The labels correspond to the most relevant quasi Lamb and Rayleigh modes. (b) Speed of sound of the calculated modes as a function of frequency. The green solid line represents the speed of sound in the saline solution.}
\end{figure}

Note that windowing artifacts appear at frequencies below 0.2 MHz and high wave numbers, thus the effect of strong scattering due to the thick porous layer (diplo\"e) in the cut-off free modes could not be reliably observed. Evidence exists of whole skull vibration modes in the audible frequency regime from experiments performed under airborne conditions \cite{McKniDBBATJotASoA2013}. However, the intermediate regime where the diplo\"e layer could play a major role is still to be explored.

The leaky skull-guided waves are further affected by scattering from inhomogeneities located outside the scanned region. The scattered waves could mimic a wave propagating faster than c0, when projected on the scanning path, thus interfering with real leaky skull-guided waves. This could explain the imperfect matching between experiments and numerical model for the leaky part of reciprocal space (Fig. 3(a)).

It is well known that leaky waves in a homogeneous plate can sustain full transmission at very specific frequencies and angles, even for large impedance mismatch between the plate and the surrounding fluid at normal incidence. Although no full  transcranial transmission is expected due to the highly heterogeneous structure of the skull and the presence of the diplo\"e, leaky skull-guided waves could be used to reach deep brain regions located at the skull's far-field.

Current far-field array and single-element approaches are not optimized to target brain regions in close proximity to the skull. Transducer arrays and focusing algorithms optimized to work at angles close to normal incidence would suffer from increased aberrations if some of the array elements face the skull close to grazing incidence \cite{Pichardo2007}. On the other hand, single element focused transducers produce an elongated focal region in the order of centimeters at the low frequencies required to reach the brain transcranially at small-angle of incidence \cite{lsombwt2014}. Thus, non-leaky skull-guided waves may present a more viable alternative to interrogate shallow brain cortex regions and the skull bone itself. 

Proper design of strategies for direct-contact excitation of leaky and non-leaky skull guided waves requires prior knowledge on the skull’s near-field properties. Our work demonstrates the existence of guided wave phenomena, thus laying ground for further studies on skull-guided wave characterization and potential applications in transcranial ultrasound.

\section{Conclusions}
Ultrasonic wave propagation in the near-field region of a water-coupled human skull has been demonstrated theoretically and experimentally for the first time, thus deepening our current understanding of ultrasound transmission by the skull bone. In addition to leaky waves, we observed non-leaky skull-guided waves corresponding to Rayleigh-Lamb waves, as predicted by a multilayer flat plate model. Further experimental and theoretical work is neccesary to fully characterize skull-guided waves, as well as to identify their excitation strategies suitable for \textit{in vivo} conditions. Observation and characterization of the skull-guided waves can be used for a more accurate interpretation of transcranial image data, such as optoacoustic images that are often afflicted by the complex wave propagation in the skull manifested via distortions in the location and shape of the vascular structures \cite{KneipTERSRJoB2016}. Our results may thus contribute to the development and optimization of non-invasive ultrasound-based techniques for diagnostic brain imaging \cite{yw2014,MeimaAGAP2017,BaykoBMNRSSAP2003,LindsNBLSUiMB2014,ShapoSWMSMIToBE2015}, monitoring of neural activity \cite{SeoNSSARCMN2016,SeoCRAMAe2013}, guided surgery applications \cite{BellOLKBP2015}, or cranial bone assesment without the use of ionizing radiation \cite{BochuVMLSR2017}.

\acknowledgments
Financial support is acknowledged from the European Research Council grant ERC-2015-CoG-682379 and the German research Foundation Grant RA1848/5-1.

\end{document}